\documentclass[
journal=jctcce,
manuscript=letter]{achemso}

\usepackage[version=3]{mhchem} % Formula subscripts using \ce{}

\usepackage{graphicx}% Include figure files
\usepackage{dcolumn}% Align table columns on decimal point
\usepackage{bm}% bold math
\usepackage{breqn}
\usepackage{subcaption}
\usepackage{color}
\usepackage{amsfonts}
\usepackage{multirow}
\usepackage{xcolor}
\usepackage{hyperref}
\usepackage{amsmath,amssymb}

\DeclareMathOperator{\erfc}{erfc}

\author{Bj\"orn Stenqvist}
\email{bjorn.stenqvist@teokem.lu.se}
\affiliation[Lund University]
{Department of Chemistry, Division of Physical Chemistry, Lund University, Sweden}
\alsoaffiliation[Lund University]
{Department of Chemistry, Division of Theoretical Chemistry, Lund University, Sweden}
\author{Vidar Aspelin}
\email{vidar.aspelin@teokem.lu.se}
\affiliation[Lund University]
{Department of Chemistry, Division of Theoretical Chemistry, Lund University, Sweden}
\author{Mikael Lund}
\email{mikael.lund@teokem.lu.se}
\affiliation[Lund University]
{Department of Chemistry, Division of Theoretical Chemistry, Lund University, Sweden}
\alsoaffiliation[LINXS]
{LINXS -- Lund Institute of Advanced Neutron and X-ray Science, Scheelev{\"a}gen 19, SE-223~70 Lund, Sweden}

\title{Generalized Moment Correction for Long-Ranged Electrostatics}
\begin{document}

\begin{abstract}
Describing long-ranged electrostatics using short-ranged pair potentials is appealing since the computational complexity scales \emph{linearly}
with the number of particles.
The foundation of this approach is to mimic the long-ranged medium response by cancelling electric multipoles within a small cutoff sphere.
We propose a rigorous and formally exact new method that cancels up to \emph{infinitely} many multipole moments and is free of operational damping parameters often required in existing theories.
Using molecular dynamics simulations of water with and without added salt, we discuss radial distribution functions, Kirkwood-Buff integrals, dielectrics, diffusion coefficients, and angular correlations in relation to existing electrostatic models.
We find that the proposed method is an efficient and accurate alternative for handling long-ranged electrostatics as compared to Ewald summation schemes.
The methodology and proposed parameterization is applicable also for dipole-dipole interactions. 
\end{abstract}

\section*{Table of Contents Graphic}
\begin{center}
\includegraphics[width=8cm]{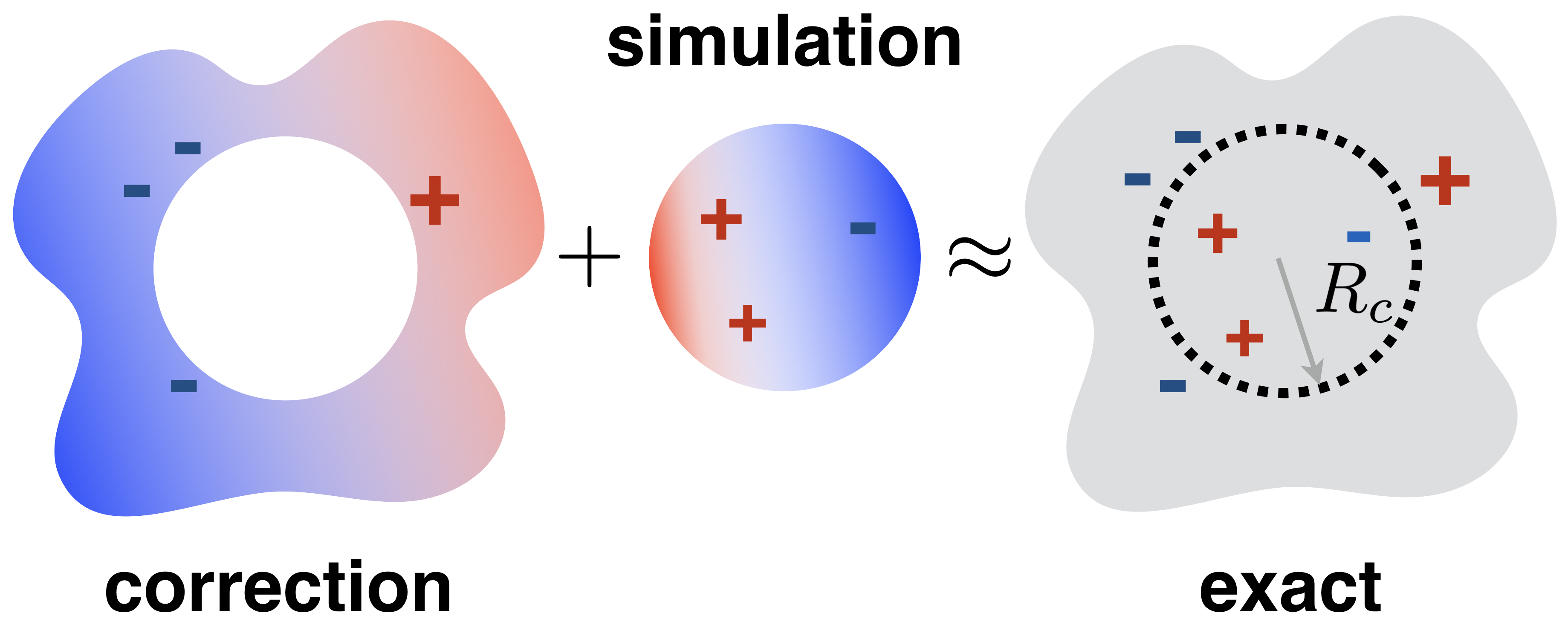}
\end{center}

\section{Introduction}
Accurate and efficient schemes to calculate long-ranged electrostatic interactions are important to expand time and space in atomistic simulations.
One of the earliest schemes is the reaction field method~\cite{Bell,Barker1973} where a spherical cutoff, $R_c$, is applied.
The potential is zero beyond the cutoff, and hence the computational complexity scales as $\mathcal{O}(\mathcal{N})$ where $\mathcal{N}$ is the number of charged particles in the system.
However, to correctly parameterize the reaction field method, the surrounding dielectric constant, $\varepsilon_{RF}$, needs to be estimated and artifacts may arise due to discontinuities at the cutoff~\cite{Stenqvist2015Direct}.

Lattice approaches, including the Ewald method\cite{ewald1921berechnung,kornfeld1924berechnung}, provides an accurate electrostatic description within a well-known parameter space\cite{kolafa1992cutoff,wang2001estimate} but are inherently limited to periodic systems.
The computational complexity of Ewald is $\mathcal{O}(\mathcal{N}^2)$ but reduces to $\mathcal{O}(\mathcal{N}^{3/2})$ with optimal parameters.
This growing cost with increasing system size makes Ewald methods demanding for large systems, albeit derived versions such as Particle Mesh Ewald~\cite{PME} (PME) has improved $\mathcal{O}(\mathcal{N}\log\mathcal{N})$ scaling.

Alternative cutoff-based methods with $\mathcal{O}(\mathcal{N})$ scaling have been shown accurate for isotropic systems and include the charge neutralizing Wolf method\cite{wolf1999exact}.
This and many derivates thereof have been thoroughly tested in the litterature~\cite{demontis2001application,zahn2002enhancement,Fennell2006,fanourgakis2009fast}.
Evolutions from the original Wolf formalism are primarily based on cancellation of the derivative of the potential at the cutoff, which is convenient since this is a prerequisite for Molecular Dynamics simulations.
Such cancellation of the derivative may be equivalent to canceling an introduced dipolar artefact\cite{Fukuda2011Molecular}.

By expanding from the Wolf method, we here show that cancellation of \emph{every} electrostatic moment -- monopole, dipole, quadrupole, octupole, etc. -- is equivalent to taking all long-ranged interactions into account in a conducting medium.
We present a potential which in addition to cancelling arbitrarily (including infinitely) many moments, cancels equally many higher derivatives of the potential at the cutoff.
From a physical view-point this is reasonable as to to avoid truncation errors; a conjecture that has previously been brought forward\cite{zahn2002enhancement}.
A similar approach of higher order moment-cancellation has recently been presented\cite{fukuda2013zero}, albeit an operational damping-parameter is introduced in addition to the cutoff distance.
The approach presented in this work requires
(i) a cutoff distance and
(ii) choosing how many electric moments, $P$, to cancel.
The integer $P$ thus connects directly to physical properties of the examined system, while avoiding the operational damping-parameter in similar formalisms.
Short-ranged potentials, as the one presented in this work, commonly assume isotropy beyond the given cutoff region, and should be cautiously used in systems where this is not a valid premise, for example at interfaces\cite{mendoza2008wolf} or in ferroelectric systems.
In the next section we derive and explore the introduced pair potential, whereafter we test it in molecular dynamics simulations of aqueous systems.

\section{Theory}

The electrostatic pair-interaction energy $u_{ij}$ between charges $z_i$ and $z_j$ can for all here examined models be written as a function of separation $r$, 
\begin{equation}
\label{eq:coulomb_splitting}
    u_{ij} = \frac{e^2z_i z_j}{4\pi\varepsilon_0\varepsilon_r r}
    \mathcal{S}(q).%, \quad r<R_c
\end{equation}
Here $e$ is the elementary charge, $\varepsilon_0$ is the vacuum permittivity, and $\varepsilon_r$ the relative permittivity of the dispersing medium. The short-ranged function, $\mathcal{S}(q)$, is defined in Table~\ref{tbl:split} using $q=r/R_c$ and $\mathcal{S}(q)\equiv 0$ for $q>1$ where the latter ensures the potential to be finite-ranged. Applying $\mathcal{S}(q)\ne 1$ results in a modified Coulomb-potential which still captures the general $1/r$ distance-dependence of the original Coulomb potential.
%$\mathcal{S}(q)$ is a modification to the original potential which thus still captures the general $1/r$ distance-dependence of a Coulomb potential. 
The new method presented here, which we denote the ``$q$-potential'', cancel an arbitrary number of electrostatic moments and derivatives of the potential at the cutoff, and is defined by the short-ranged function
\begin{equation}\label{eq:qpot}
    \mathcal{S}(q) = \prod_{n=1}^{P}(1-q^n).
\end{equation}
%In the next section we will show the theory on which it is based and how it was derived. 
In the next section we formally derive the theory.

\begin{table}[ht]
    \centering
    \begin{tabular}{lll}\hline
     label         & $\mathcal{S}(q)$        & ref.\\\hline
     Coulomb & 1 & \\
     Reaction-field & $1+q^3\left(\frac{\varepsilon_{RF}-1}{2\varepsilon_{RF}+1}\right)$ & \cite{Barker1973}\\
     Ewald, real space & $ \erfc{(\eta q)} $ & \cite{ewald1921berechnung}\\
     SP0 (Wolf)          & $ \erfc{(\eta q)}-q\erfc{(\eta)} $ & \cite{wolf1999exact}\\
     SP1        & $(1-q)^2$ & \cite{Fennell2006} \\
     SP3   & $(1 + 2.25q + 3q^2+2.5q^3)(1-q)^4$ & \cite{fanourgakis2009fast}\\\hline
     $q$-potential   & $\prod_{n=1}^{P}(1-q^n)$ & this work\\\hline
    \end{tabular}
    \caption{Electrostatic schemes and their short-ranged functions, $\mathcal{S}(q)$.
    $\eta = \alpha R_c$ where $\alpha$ is a damping-parameter. SP means ``shifted potential'' and the suffix denotes the number of potential derivatives that are zero at the cutoff.}
    \label{tbl:split}
\end{table}

\subsection{Generalized Moment Cancellation}\label{sec:gentheory}

%A fair approximation to the effective Coulomb pair potential can be achieved by neutralizing the total charge within a cutoff with an oppositely charged distribution on the surface of the cutoff-region\cite{Wolf}. By further cancel the total dipole moment the potential depends less on the damping-parameter $\kappa$ introduced in the previous formalisms\cite{Wolf,Yonezawa2012}. Following this trend, we here now set out to cancel \emph{arbitrarily} many moments and explain why this is reasonable to do.

\begin{figure}[!ht]
\centering
   \includegraphics[width=0.5\columnwidth]{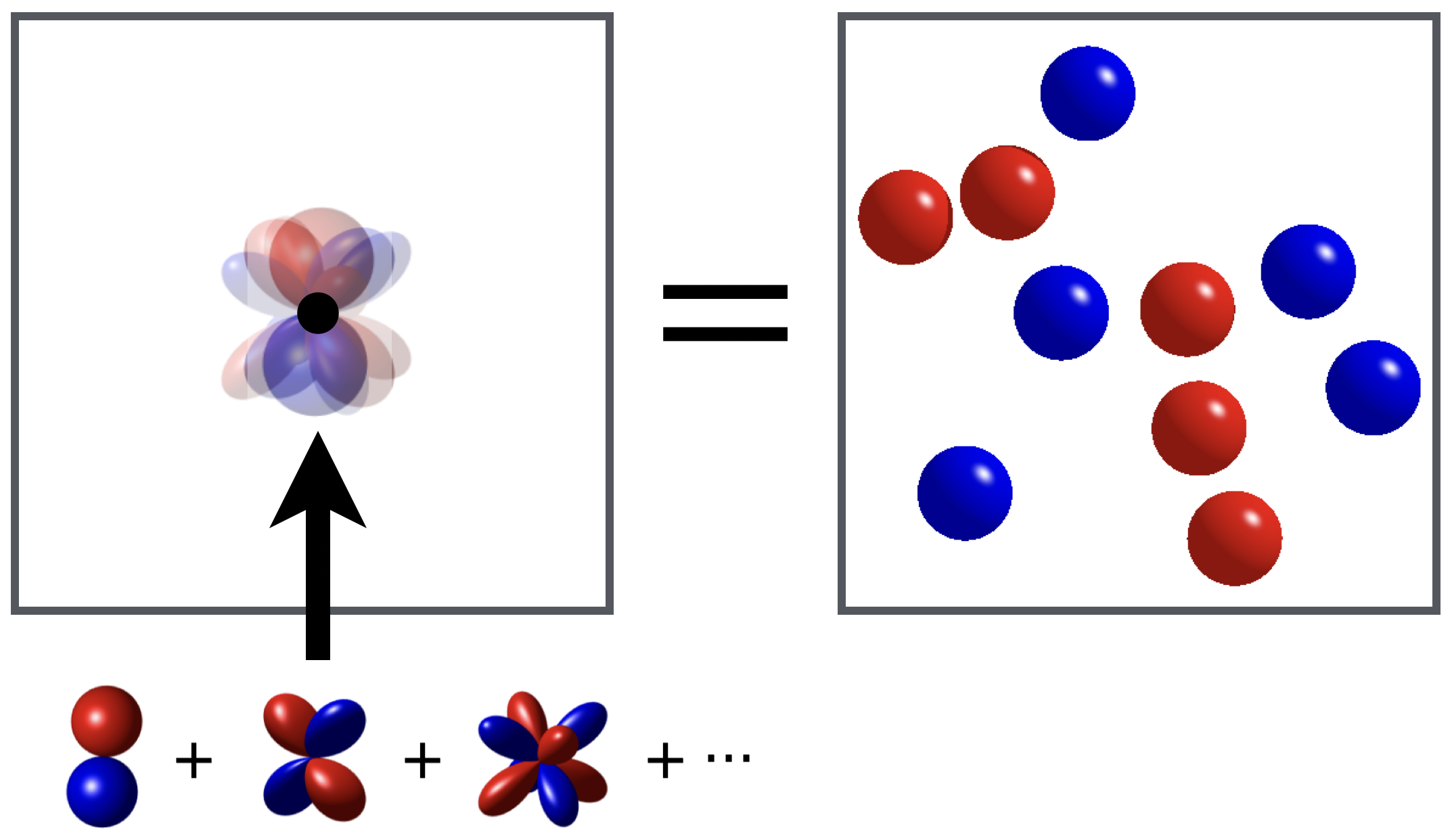}
\caption{A single point-multipole particle (left) describing a point-charge distribution (right).}
\label{fig:box_conv} 
\end{figure}

The total interaction energy between $\mathcal{N}$ charged particles using periodic boundary conditions (PBC) is
\begin{equation}
\label{eq:energy_pbc}
E_{\rm Tot} = \frac{e^2}{4\pi\varepsilon_0\varepsilon_r}\frac{1}{2}\sum_{\boldsymbol{n}\in \mathbb{Z}^3}^{\prime}\sum_{i=1}^{\mathcal{N}}\sum_{j = 1}^{\mathcal{N}} z_i\boldsymbol{T}_0\left( \boldsymbol{r}_{ij} + \boldsymbol{n}\circ\boldsymbol{L} \right) z_j.
\end{equation}
Here the prime indicates that $i\ne j$ when $\boldsymbol{n}=\boldsymbol{0}$, $\circ$ denotes the Hadamard product, $\boldsymbol{r}_{ij}$ is the distance-vector between point-charges $i$ and $j$, the size of the cuboid cell is described by its side-lengths $\boldsymbol{L}=(L_x,L_y,L_z)$, and the zeroth order interaction-tensor $\boldsymbol{T}_0(\boldsymbol{r})$ is
\begin{equation}
\boldsymbol{T}_0(\boldsymbol{r}) = \frac{1}{|\boldsymbol{r}|}.
\end{equation}
By assuming $|\boldsymbol{L}|/2 < \min(\boldsymbol{L})$ it is possible to convert all point-charge particles in each replicated cell into a single centered point-multipole particle represented by infinitely many higher order moments\cite{bottcher1978theory,stone2013theory} as in Fig.~\ref{fig:box_conv}. This conversion is valid as the center of any reciprocal point-multipole particle, positioned at $\boldsymbol{n}\circ\boldsymbol{L}$, is located further away from the origin of the centered cell ($\min_{\boldsymbol{n}}(|\boldsymbol{n}\circ\boldsymbol{L}|)\ge \min(\boldsymbol{L})$) than any point-charge particle in the centered cell itself ($\max(|{\bf r}_{i}|)=|\boldsymbol{L}|/2$ where ${\bf r}_{i}$ is the distance-vector from point-charge $i$ to the origin). 
If every moment of every such point-multipole particle is a zero-tensor, then there would not be any interactions between reciprocal cells, \emph{i.e.} no long-ranged interactions. For such a system the interaction-energy would simply be
\begin{equation}
\label{eq:energy_final}
E_{\rm Tot} = \frac{e^2}{4\pi\varepsilon_0\varepsilon_r}\frac{1}{2}\sum_{i=1}^{\mathcal{N}}\sum_{\substack{j = 1 \\ j \ne i }}^{\mathcal{N}} z_i\boldsymbol{T}_0(\boldsymbol{r}_{ij}) z_j.
\end{equation}
Ergo, if an effective potential cancels all the electrostatic moments of the cell then the total interaction energy would be given solely from the particles in the centered box. Based on the above description the procedure can be regarded as a reaction field approach where the induced field cancels moment locally.
Even if elements in higher order tensors does not fully cancel and therefore are not exactly zero, Eq.~\ref{eq:energy_final} might be a highly accurate approximation when the short-ranged interactions (described by higher order tensors) from reciprocal cells only gives minor contributions. Consequently, even imperfect conductors with finite $P$ can give accurate results for the long-ranged electrostatics. In fact, for imperfect conductors like water, an infinite $P$ might cancel too many interactions and thus give erroneous results. For higher order interactions than dipole-dipole and ion-quadrupole interactions, the long-ranged interactions are absolutely convergent. Therefore, the highest order moment that in theory needs to be cancelled, for sufficiently large systems, is the quadrupole moment, \emph{i.e.} $P=3$ which we also see in Section~\ref{sec:results}. The same arguments as explained in this section hold for interactions between any regions with zero or small tensor moments and the cancellation approach is therefore not limited to PBC systems.

\subsection{Derivation of the $q$-potential}\label{app:A}
\begin{figure}[ht]
\centering
   \includegraphics[width=0.8\columnwidth]{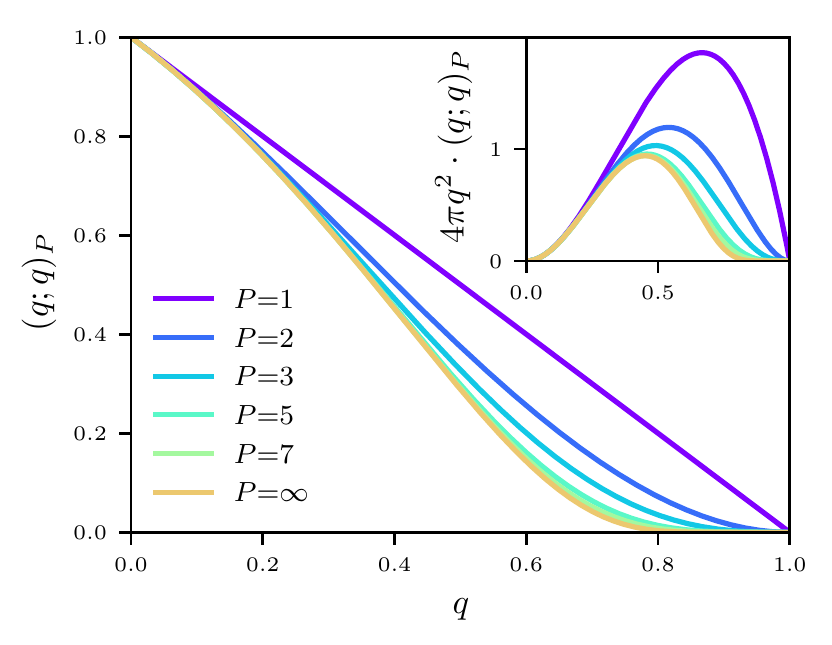}
\caption{The short-ranged function $\mathcal{S}(q)=(q;q)_P=\prod_{n=1}^P(1-q^n)$ for different numbers of cancelled moments, and (inset) the same scaled with the Jacobian $4\pi q^2$ as to assess its volumetric influence in calculations. }
\label{fig:Pochhammer} 
\end{figure}

We begin the derivation by assuming that an electrostatic potential based on moment cancellation can be described by the potential from the original particle and $P$ image particles. The aggregated potential from the original and all image particles is denoted $V_q$, which is a function of the distance-vector ${\bf r}$ and the charge $z$,
\begin{equation}
\label{eq:imag_pot}
V_{q}({\bf r},z) = V({\bf r},z) + \sum_{p=1}^P V({\bf r}_p,z_p).
\end{equation}
Here $V$ represents the Coulomb potential,
$z$ the charge of the original particle,
$z_p$ the charge of image particle $p$, and ${\bf r}_p = c_p{\bf r}$ where $c_p$ is a proportionality factor to be discussed.
The requirement that up to $M$ higher order moments of all image particles should cancel the original particle moments, can be formulated as
\begin{equation}
\label{eq:finalMatrix}
\begin{bmatrix}
       1    \\[0.3em]
       r  \\[0.3em]
       \vdots          \\[0.3em]
       r^{M}      
     \end{bmatrix}z + \begin{bmatrix}  
  1 & 1 & \cdots & 1 \\
  r_1 & r_2 & \cdots & r_P \\
  \vdots  & \vdots  & \ddots & \vdots  \\
  r_1^{M} & r_2^{M} & \cdots & r_P^{M}
     \end{bmatrix}
     \begin{bmatrix}
       z_{1}    \\[0.3em]
       z_{2}  \\[0.3em]
       \vdots          \\[0.3em]
       z_{P} 
     \end{bmatrix}
     = \begin{bmatrix}
       0    \\[0.3em]
       0  \\[0.3em]
       \vdots          \\[0.3em]
       0      
     \end{bmatrix}.
\end{equation}
Here $r$ is any component of ${\bf r}$, and $r_p$ is the corresponding component of ${\bf r}_p$. The meaning of row $m$ in Eq.~\ref{eq:finalMatrix} is that the $m-1$ order moment from the original particle (first term; $r^{m-1}z$) together with the sum of the same order moments of the image particles (second term; $\sum_{p=1}^Pr_p^{m-1}z_p$) should equal zero (right-hand-side).
Assuming that the image particle positions $r_p$ are unique and non-zero, and that the matrix is square ($M=P-1$), we ensure that a solution exists.\cite{Macon1958}
By then choosing the positions $r_p$ we can extract the charges, $z_p$.
The positions of the image particles may be arbitrary chosen, with the exceptions mentioned earlier, however in this work we use $c_p = q^{-p}$.
That is, the position of image particle $p+1$ ($r_{p+1}$) is the mirror image of the position of image particle $p-1$ ($r_{p-1}$) in the position of image particle $p$ ($r_{p}$), \emph{i.e.} $r_p^2 = q^{-2p}r^2 = q^{-(p-1)}rq^{-(p+1)}r = r_{p-1}r_{p+1}$.
This choice is closely related to the method of image charges\cite{tikhonov1963equations,friedman1975image}.
Together with Eq.~\ref{eq:finalMatrix}, the above assumptions yield the solution for the image charges as
\begin{equation}
\label{eq:solutionH}
z_{p} = z\cdot {P \brack p}_q(-1)^pq^{p(p-1)/2}
\end{equation}
where ${P \brack p}_q $ is the \emph{q}-analogue of the binomial coefficients\cite{comtetadvanced,kac2002quantum}.
For details see the supplemental information (SI), Section~S1.
By using Eq.~\ref{eq:solutionH} it is further possible to present an expression for the aggregated potential $V_{q}$, and the modified interaction-tensor as
\begin{equation}
\boldsymbol{T}^{q}_0(\boldsymbol{r}) = \boldsymbol{T}_0(\boldsymbol{r})\left(1 +  \sum_{p=1}^{P}\frac{{P \brack p}_q(-1)^pq^{p(p-1)/2}}{q^{-p}}\right).
\end{equation}
The expression within the parenthesis is composed of the contributions from the original particle (the $1$) and all image particles (the sum).
Rearranging and simplifying\cite{kac2002quantum} this expression gives
\begin{equation}
\label{eq:modified_tensor}
\boldsymbol{T}^{q}_0(\boldsymbol{r}) = \boldsymbol{T}_0(\boldsymbol{r})\prod_{n=1}^{P}(1-q^n) = \boldsymbol{T}_0(\boldsymbol{r})(q;q)_{P}
\end{equation}
where $(a;q)_{P}$ is the \emph{q}-Pochhammer Symbol. In Fig.~\ref{fig:Pochhammer} we show this multiplicative modification to the original interaction-tensor for different $P$. Since the multiplicative term is a $q$-analogue of the Pochhammer symbol we address the proposed potential as the \emph{q}-potential.

By re-writing Eq.~\ref{eq:modified_tensor} as
\begin{equation}
\boldsymbol{T}^{q}_0(\boldsymbol{r}) = \boldsymbol{T}_0(\boldsymbol{r})(1-q)^P\prod_{n=1}^{P}\sum_{k=0}^{n-1}q^k
\end{equation}
and specifically note the $(1-q)^P$-term, we acknowledge that the number of cancelled \emph{higher} order moments, $P-1$, equals the number of \emph{higher} order derivatives with respect to $|\boldsymbol{r}|$ (or equivalently $q$) that are zero at the cutoff.
Finally, using the already described theory, but with image particle positions $c_p=q^{-sp}$ we retrieve the short-ranged function
\begin{equation}\label{eq:final_exp}
\boldsymbol{T}^{q}_0(\boldsymbol{r}) = \boldsymbol{T}_0(\boldsymbol{r})(q^s;q^s)_{P}
\end{equation}
where $s\in\{1,2,...\}$, $s-1$ higher order derivatives are cancelled also at $q=0$. This is straight-forward to acknowledge as the lowest power of $q$ in $(q^s;q^s)_{P}$ (save $q^0=1$) is $q^s$. Consequently, the $(s-1)$th derivative is the highest order derivative to vanish at $q=0$ (given that $P>0$).
The expression in Eq.~\ref{eq:final_exp} can now be tuned in how many derivatives to cancel at $q=0$ (using $s$) and at $q=1$ (using $P$).
Due to the physical interpretation of the image charge positions\cite{tikhonov1963equations,friedman1975image} we nonetheless in this work use $s=1$.

A pair-interaction entails two particles and above we cancelled moments of one of them, specifically the one at distance $r$ from the origin (where the other particle is located). The charge of the origin-located particle has thus not been cancelled, yet we now address this issue in a similar manner as the previously explained moment cancellation procedure. The resulting energy is defined as the \emph{self energy}, and in the SI Section~S2 the detailed derivation is presented which culminates into
\begin{equation}
\label{eq:self_energy}
E_{{\rm Self}} = -\frac{e^2}{4\pi\varepsilon_0\varepsilon_rR_c}\sum_{i = 1}^{\mathcal{N}} z_i^2.
\end{equation}

\subsection{Potential for systems with moments}

The no net-moment assumption is fair for many liquids, but generally not so at interfaces or in systems with dipolar or ferroelectric properties.
For such systems we suggest a generalization of the presented methodology.
We previsously noted that to cancel every moment within a cutoff region gives the right-hand side of Eq.~\ref{eq:finalMatrix} as the zero-vector.
If, however, there would be a non-zero moment within this region, one gets Eq.~\ref{eq:finalMatrix_mom} where $\Psi^{(p)}$ is the $p$th higher order moment in the region projected onto the $r$-component of ${\bf r}$.
The solution to this equation is more comprehensive than the solution to Eq.~\ref{eq:finalMatrix}, yet solvable, at least given the mentioned restrictions on $r_p$ and $M$.
Here we have not explicitly expanded the given framework for such cases, yet we highlight the possibility to develop a $q$-potential for non-zero moment environments.

\begin{equation}
\label{eq:finalMatrix_mom}
\begin{bmatrix}
       1    \\[0.3em]
       r  \\[0.3em]
       \vdots          \\[0.3em]
       r^{M}      
     \end{bmatrix}z + \begin{bmatrix}  
  1 & 1 & \cdots & 1 \\
  r_1 & r_2 & \cdots & r_P \\
  \vdots  & \vdots  & \ddots & \vdots  \\
  r_1^{M} & r_2^{M} & \cdots & r_P^{M}
     \end{bmatrix}
     \begin{bmatrix}
       z_{1}    \\[0.3em]
       z_{2}  \\[0.3em]
       \vdots          \\[0.3em]
       z_{P} 
     \end{bmatrix}
     = \begin{bmatrix}
       \Psi^{(0)}    \\[0.3em]
       \Psi^{(1)}  \\[0.3em]
       \vdots          \\[0.3em]
       \Psi^{(M)}      
     \end{bmatrix}
\end{equation}

\subsection{Choosing the cutoff}
\label{sec:cutoff}
It is desirable to use a minimal cutoff to speed up calculation time. The physical interpretation of such a cutoff which still accurately describes the system is: the smallest region which exhibits small enough fluctuations in the aggregated moments to be corrected for by induced image particles. It is formally sound to use a larger cutoff as larger regions maintain this quality.
Yet, most systems do display a certain local anisotropy, or equivalently, non-vanishing higher order moments.
Accordingly, the cutoff needs to enclose this whole space. To ensure no inter-cell correlations between anisotropic regions, it should not exceed one fourth of the shortest cell-length\cite{stenqvist2018replicate}.
Finally, the no net-moment approach does not impose an isotropic cutoff and any closed space shape can be used. The choice of cutoff shape, we speculate, may be particularly important in strongly anisotropic regions such as interfaces.

\section{Methods}
\subsection{Molecular Dynamics Simulation}
All simulations were performed in the isobaric-isothermal ensemble at pressure $\mathcal{P}=$1~bar and temperature $T=$298.15~K, using OpenMM 7~\cite{Eastman2017}. The pressure was kept constant using a Monte Carlo barostat\cite{qvist2004} and we used a Langevin integrator~\cite{Sivak2014}, with a friction coefficient of 1.0~ps and a time step of 2.0~fs, to keep constant temperature. Before production runs, the systems were energy minimized, then equilibrated. The reference Ewald summation/PME simulations used a fractional force error tolerance of $5\cdot 10^{-4}$. 

For simulations of only SPC/E water molecules, Ewald summation was used as reference, $\mathcal{N}=2000$ and the production runs spanned 10~ns. Three different atom-based (real space) cutoff distances were used for all electrostatic schemes: $0.96$~nm, $1.28$~nm, and $1.60$~nm, which roughly corresponds to $3$, $4$, and $5$ oxygen ($\sim$water) diameters. The Lennard-Jones interactions used the same cutoff and a long-ranged correction term to compensate for the neglected contributions\cite{Shirts2007Accurate}.
Systems containing SPC/E water molecules and NaCl/NaI ions were simulated to retrieve activity derivatives according to Kirkwood-Buff theory as described in the next subsection. The simulations were carried out using $R_c=1.28$~nm in a cubic box with $\mathcal{N}=4402-5963$ water molecules depending on the type and concentration of electrolyte presented in Appendix~\ref{app:F}. The production runs spanned 40~ns and PME was used as reference. 

The diffusion coefficient defined in Eq.~\ref{eq:diffcoeff} was calculated by least-square fitting of the mean-square displacement (MSD) of the particles throughout the simulation to a linear curve, where the slope is the sought after value. The MSD was sampled in the micro-canonical ensemble, which continued upon the last state of the above explained isobaric-isothermal simulations with a 1~ns equilibration and 2~ns production run. The micro-canonical production simulation also provided the base for the analysis of the energy.
\begin{equation}\label{eq:diffcoeff}
D = \lim_{t\to\infty}\left(\frac{1}{6}\frac{{\text d}\langle | \boldsymbol{r}(t)-\boldsymbol{r}(0) |^2 \rangle}{{\text d}t} \right)
\end{equation}

The standard deviations were based on samples from ten consecutive, equally sized intervals that spanned the whole simulation. 

\subsection{Kirkwood-Buff Theory}

The Kirkwood-Buff (KB) theory provides a general way to obtain macroscopic properties of a solution from its microscopic properties \cite{kirkwood1951statistical}. The central property is the Kirkwood-Buff integral (KBI) between components $A$ and $B$, defined as
\begin{equation}
G_{AB} = 4\pi\int_{0}^{\infty}[g_{AB}^{\mu VT}(r)-1]r^2\textrm{d}r
\label{eq:defKBint_red}
\end{equation}
where $g_{AB}^{\mu VT}(r)$ is the radial distribution function (RDF) in the grand canonical ensemble (\emph{i.e.} an open system), and $r$ is the distance between the components. To obtain approximate KBIs in a closed system, the integral is typically truncated at a distance $R$ after which the RDFs converge, and the grand-canonical RDF is replaced by the RDF computed in the closed system \cite{Weerasinghe2003, Hess2009}. To obtain accurate approximate KBIs according to this procedure, corrections are needed to compensate for the introduced errors. We followed a procedure\cite{Milzetti2018} in which two corrections\cite{Ganguly2013,Krger2013} are applied simultaneously to the KBIs. These correction factors are further explained in Appendix~\ref{app:F}. The derivative of the electrolyte activity at constant pressure, and temperature, is obtained according to \cite{BenNaim1992}
\begin{equation}
a_{c}' = \left.\frac{\partial \ln a_c}{\partial \ln \rho_c}\right |_{\mathcal{P}T}=\frac{1}{1+\rho_{c}(G_{cc}-G_{wc})}.
\label{eq:actDer}
\end{equation}
Here $G_{cc}$ and $G_{wc}$ are the corresponding truncated, corrected KBIs obtained from simulations, $\hat{G}^*_{cc}(R)$ and $\hat{G}^*_{wc}(R)$ respectively (see Appendix~\ref{app:F}). Subscripts $c$ and $w$ denote electrolyte and water respectively, $a_c = \gamma_c \rho_c$ is the electrolyte activity, $\gamma_c$ is the molar mean activity coefficient of the electrolyte, and $\rho_c$ is the number density of the electrolyte. For the water-electrolyte KBI, we chose to use the RDF between the water oxygen and the electrolyte, in the following denoted $g_{Oc}(r)$. Experimental activity derivatives for the simulated electrolytes were calculated from previously reported activity coefficients\cite{Robinson1959}, using the fitting procedure described in Appendix~\ref{app:F}. For the simulations, we used one force field\cite{Tesei2018} for the Na$^+$ and I$^-$ ions whilst another\cite{Dang1995} was used for Cl$^-$. The parameters of these force fields are presented in Appendix~\ref{app:F}.

\section{Results}\label{sec:results}
To evaluate the developed $q$-potential we investigate a bulk water system as well as an aqueous salt solution by analyzing, among others, radial distributions functions, angular correlations, and KBIs.
While the Ewald/PME methods assume a replicated environment and may therefore not necessarily represent a true isotropic system\cite{hunenberger1999effect,stenqvist2018replicate}, we choose this as a reference system due to its widespread use in molecular simulations.

\subsection{Bulk water-system}\label{sec:bulk}
\begin{figure}[!t]
    \centering
    \includegraphics[width=1.0\columnwidth]{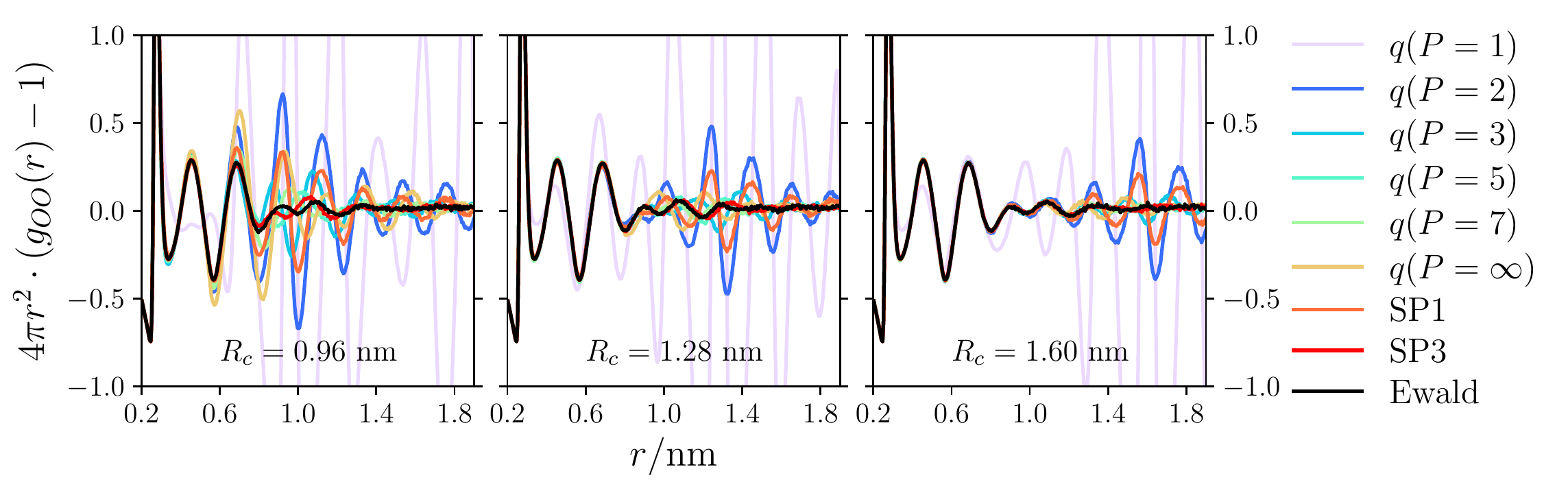}
    \caption{The deviation from bulk presented as $g_{OO}(r)-1$ scaled by the Jacobian $4\pi r^2$, where $g_{OO}(r)$ is the oxygen-oxygen RDF. For clarity we have faded the strongly oscillating $q(P=1)$ lines.}
    \label{fig:rdf_m1}
\end{figure}

A representation of the radial distribution functions between water oxygen atoms are presented in Fig.~\ref{fig:rdf_m1}, where the $q(P=1)$-potential stands out with a clear peak at the respective cutoff distance. By initially increasing the order $P$ from $2$, we get results closer resembling Ewald. However, by using $q(P=\infty)$ and $R_c\le 1.28$~nm we diverge from the Ewald-like results retrieved by intermediate $P$-values. 
The SP1-approach cancels one derivative at the cutoff and yields results more closely resembling Ewald than the similar $q(P=2)$-potential. Yet, the $q(P=3)$-potential yields results even closer to Ewald. The SP3-approach, which cancels three derivatives at the cutoff, produces among the most Ewald-like results akin the $q(P=5)$- and $q(P=7)$-potentials.

\begin{figure}[!t]
    \centering
    \includegraphics[width=1.0\columnwidth]{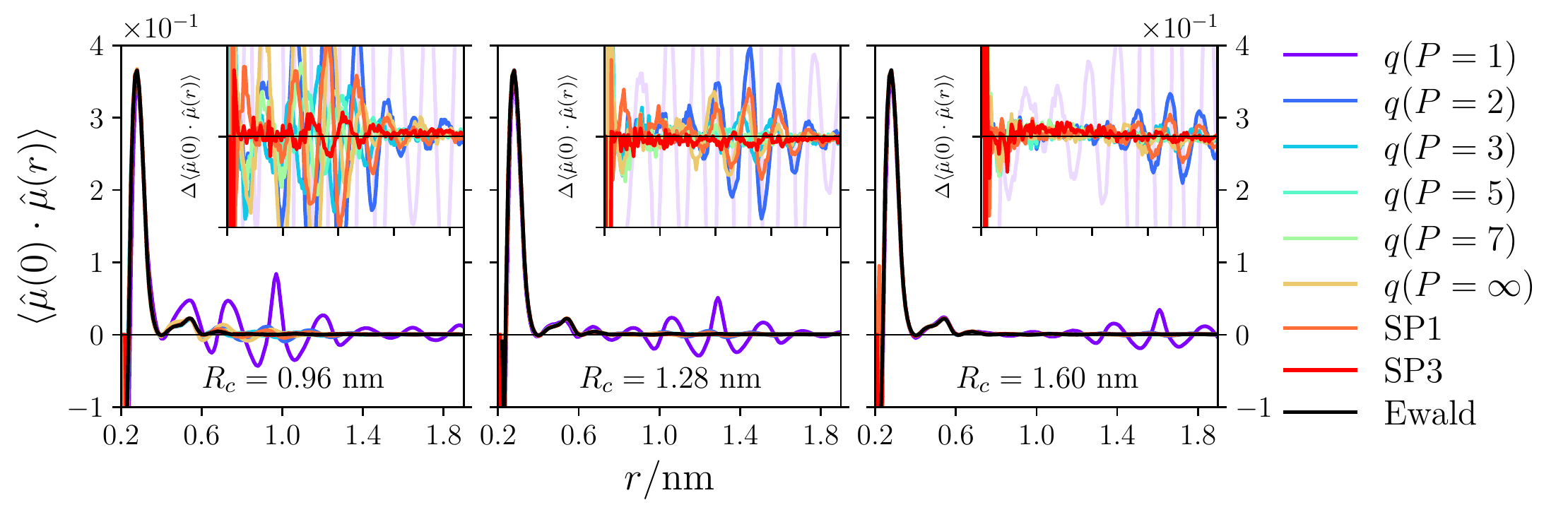}
    \caption{Dipole-dipole correlation using normalized dipole $\hat{\mu}$, and (inset) the difference to Ewald where the ordinate spans $\pm 5\cdot 10^{-3}$.}
    \label{fig:mumu}
\end{figure}

\begin{table}[ht]
    \centering
    \begin{tabular}{ c | c c c | c c | c c c | c c | c c c | c c |}
    %\hline
    \multirow{2}{*}{Potential} & \multicolumn{5}{c|}{$R_c=0.96$~nm} & \multicolumn{5}{c|}{$R_c=1.28$~nm} & \multicolumn{5}{c|}{$R_c=1.60$~nm} \\ & $\rho$ & $\varepsilon_r$ & $G_K$ & $\sigma_E$ & $D$ & $\rho$ & $\varepsilon_r$ & $G_K$ & $\sigma_E$ & $D$ & $\rho$ & $\varepsilon_r$ & $G_K$ & $\sigma_E$ & $D$ \\\hline
    $q(P=1)$ & 1137 & --- & 4.8 & --- & --- & 1106 & ---  & 3.2 & --- & --- & 1073 & --- & 3.0 & --- & --- \\
    $q(P=2)$ & 1005 & 78 & 3.0 & 24 & 2.2 & 1002 & 75 & 3.0 & 15 & 2.2 & 1000 & 69 & 2.9 & 11 & 2.7 \\
    $q(P=3)$ & 1002 & 67 & 2.9 & 10 & 2.3 & 1001 & 70 & 3.0 & 8 & 2.5 & 1000 & 76 & 3.0 & 3 & 2.4 \\
    $q(P=4)$ & 1002 & 67 & 2.9 & 9 & 2.4 & 1000 & 67 & 2.9 & 6 & 2.5 & 1000 & 69 & 2.9 & 10 & 2.4 \\
    $q(P=5)$ & 1003 & 71 & 3.0 & 8 & 2.2 & 1001 & 67 & 2.9 & 8 & 2.6 & 1000 & 72 & 2.9 & 12 & 2.3 \\
    $q(P=6)$ & 1004 & 76 & 3.2 & 5 & 2.4 & 1001 & 71 & 3.0 & 4 & 2.4 & 1000 & 68 & 2.9 & 3 & 2.7 \\
    $q(P=7)$ & 1005 & 73 & 3.0 & 8 & 2.5 & 1001 & 68 & 2.9 & 4 & 2.4 & 1000 & 69 & 2.9 & 3 & 2.5 \\
    $q(P=8)$ & 1006 & 72 & 3.0 & 9 & 2.5 & 1001 & 70 & 2.9 & 6 & 2.6 & 1000 & 71 & 3.0 & 6 & 2.5 \\
    $q(P=\infty)$ & 1008 & 73 & 3.1 & 9 & 2.4 & 1001 & 66 & 2.9 & 4 & 2.7 & 1000 & 67 & 2.9 & 4 & 2.5 \\\hline
    SP1 &   993 & 67 & 2.9 & 1 & 2.9 &   996 & 69 & 2.9 & 1 & 2.8 &   996 & 68 & 2.9 & 1 & 2.8 \\
    SP3 &   998 & 73 & 3.0 & 0 & 2.4 &   998 & 66 & 2.9 & 0 & 2.5 &   998 & 71 & 3.0 & 0 & 2.5 \\\hline
    Ewald & 998 & 67 & 2.9 & 4 & 2.5 & 998 & 73 & 3.0 & 2 & 2.6 & 998 & 69 & 3.0 & 6 & 2.9 \\
    \hline
    Exp. & 997 & 79 & --- & --- & 2.3 & 997 & 79 & --- & --- & 2.3 & 997 & 79 & --- & --- & 2.3 \\
    %\hline
    \end{tabular}
    \caption{Density $\rho$ [kg/m$^3$], relative dielectric constant $\varepsilon_r$ [unitless], Kirkwood factor $G_K$ [unitless], standard deviation of total energy $\sigma_E$ [kJ/mol], and diffusion coefficient $D$ [m$^2$/s/10$^{-9}$], for the different potentials applied on a bulk water-system and experimental reference\cite{harned1958physical,Mills1973Self}.
    The standard deviations for the density, dielectric constant, and diffusion coefficient are: $\sigma_{\rho} \sim 2$ kg/m$^3$, $\sigma_{\varepsilon_r} \sim 1$ (over last half of simulation), and $\sigma_D \sim 0.1$. The exceptions are all for $P=1$ which gives higher values.}
    \label{tbl:dente}
\end{table}

The density, dielectric constant (see Section~S3 in the SI), Kirkwood factor $G_K$ (see Eq.~\ref{eq:kirkwood}), standard deviation of the total energy, and diffusion coefficient, for the different potentials are presented in Table~\ref{tbl:dente}. Again we note that $q(P=1)$ gives distinctly different results than the others and will therefore not be commented from here on.
The densities for all other $q$-potentials are generally slightly above a thousand. However, with a standard deviation of $\sim 2$, nearly all potentials (for $R_c\ge 1.28$~nm) are statistically indistinguishable.
%In contrast, the SP1 gives a lower density than Ewald, with a relative deviation similar to that between the $q$-potential and Ewald.
%SP3 gives a very similar density as Ewald and experimental values. 
%Some discrepancy between truncated potentials and Ewald are to be expected and in this case reflects stronger ($q$-potential) or weaker (SP1) electrostatic interactions compared to how the force field was originally developed.

Although the dielectric constant is seen to be irregular in $P$, the results using $q(P>1)$, SP1, and SP3 are in reasonable agreement with Ewald and experimental values. For the Kirkwood factor, all potentials with $P\ne 1$ are consistent within the range $3.0\pm 0.1$, except $q(P=6)$ which for $R_c=0.96$~nm gives $G_K = 3.2$. When comparing the standard deviation of the energies of the pair potentials to the, for PBC formally exact, Ewald reference we observe low values for all approaches. Especially SP1 and SP3 yields particularly low values, even compared to Ewald. Yet, a seemingly odd behavior is the increase in $\sigma_E$ for $P=4$ and $P=5$ while expanding the cutoff from $R_c=1.28$~nm to $R_c=1.60$~nm. A similar trend is also observed for the Ewald summation result. The diffusion coefficients are all similar to results of previous studies\cite{Spoel1998systematic,Pekka2001Structure} with $D=2.5\pm 0.3$ for all pair potentials with $R_c\ge 1.28$~nm.

Finally as stated in Section~\ref{sec:cutoff} we note that the cutoff should obey $R_c<\min(L_x,L_y,L_z)/4$ and therefore, since the box-length is $\sim 3.9$~nm, artifacts may be present in systems using excessive cutoffs. However, in the SI Section~S4 we present simulation results based on larger systems and see no effects not previously pointed out.%, such as a system size-dependent diffusion coefficient.

\begin{equation}
\label{eq:kirkwood}
    G_K = \left\langle \sum_{i=1}^{\mathcal{N}}\sum_{j=1}^{\mathcal{N}}\cos(  \hat{\mu}_i\cdot \hat{\mu}_j )\right\rangle
\end{equation}

\subsection{Salt solutions}

\begin{figure}[!ht]
    \centering
    \includegraphics[width=0.9\columnwidth]{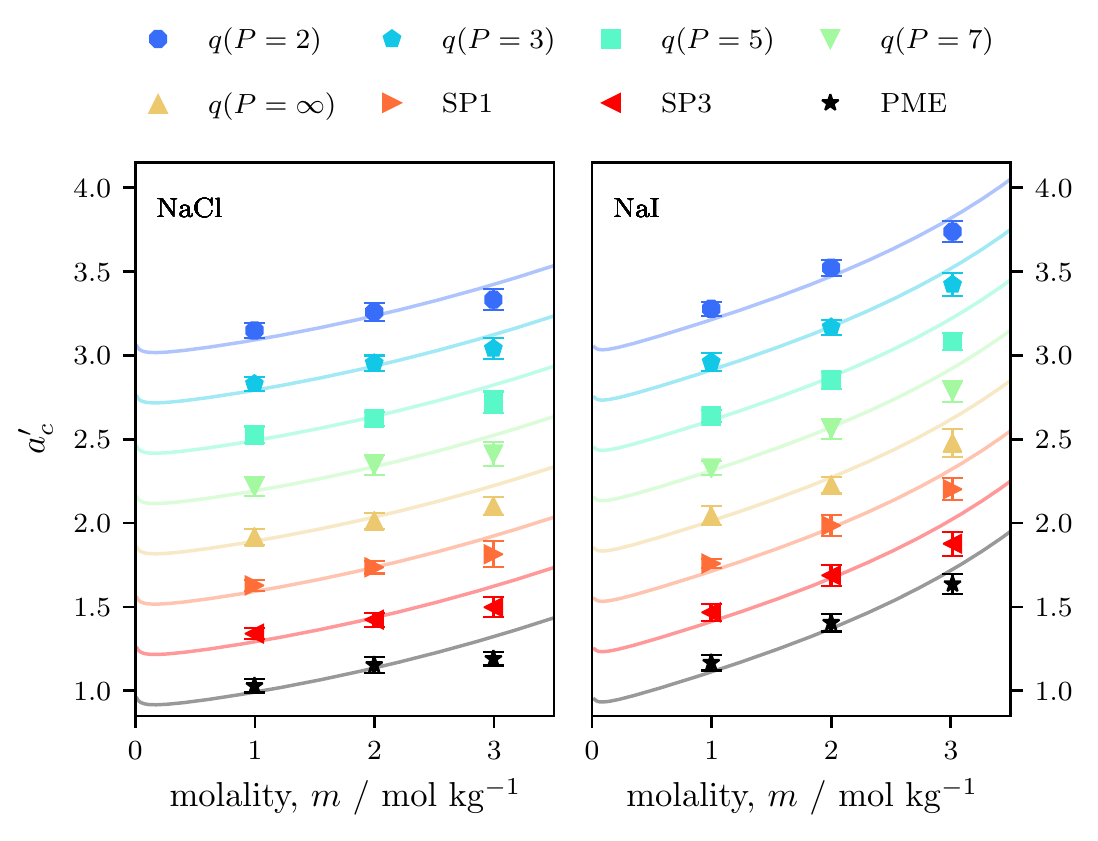}
    \caption{Activity derivatives for NaCl (left) and NaI (right) using different pair potentials, with $R_c=1.28$ nm. Solid curves represent experimental data.\cite{Robinson1959, Gee2011} For visual clarity, the PME results are plotted on top of the actual experimental curves (in black) whereas the results from the other pair potentials are shifted in steps of $0.3$ and plotted on top of the correspondingly shifted experimental curves.}
    \label{fig:actDers}
\end{figure}

Fig.~\ref{fig:actDers} shows activity derivatives for NaCl and NaI solutions and the $q(P=3)$-potential generally gives the most accurate results as compared to PME. Yet, the presented error-bars for all methods do overlap which indicates comparable results. We have performed simulations using the $q(P=1)$-potential with poor outcome (akin for the RDF), wherefore we conclude cancellation larger than one is needed to capture the nature of the system. For $m = 3$ mol kg$^{-1}$, all potentials - including PME - yield systematically lower activity derivatives compared to experimental data.

\begin{figure}[!ht]
    \centering
    \includegraphics[width=0.9\columnwidth]{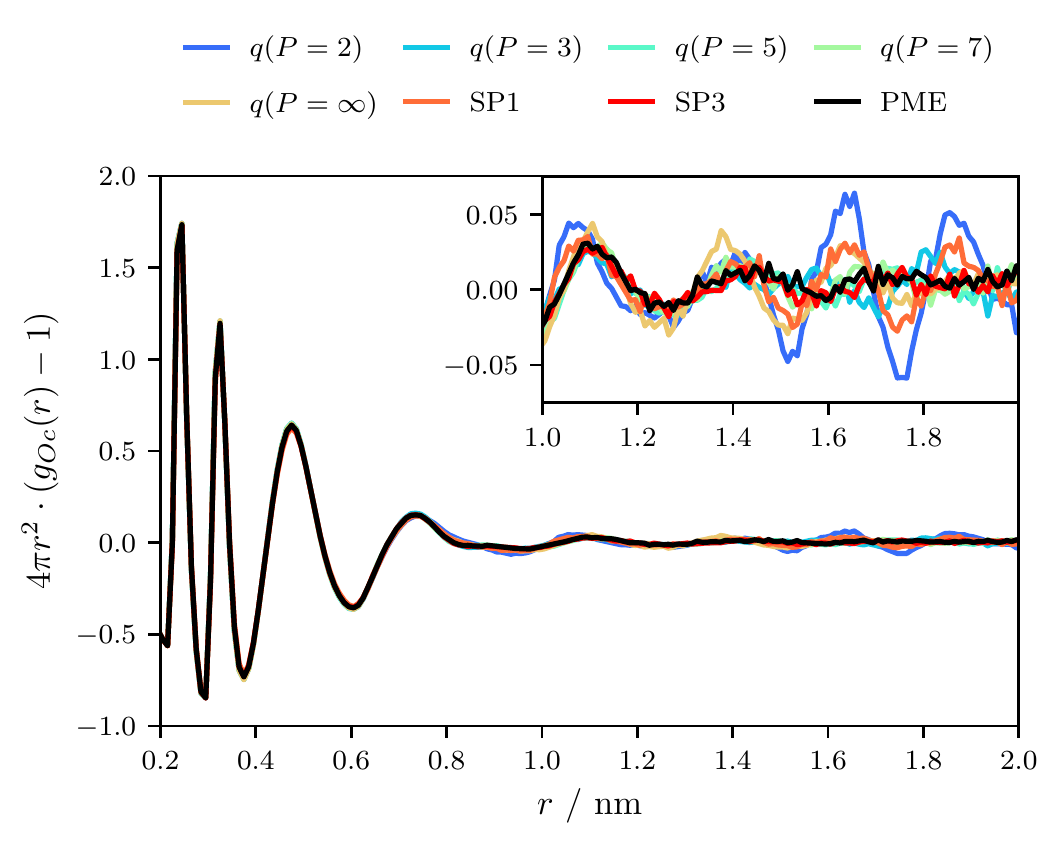}
    \caption{The deviation from bulk plotted as $g_{Oc}(r)-1$ scaled with the Jacobian $4\pi r^2$, where $g_{Oc}(r)$ is the water oxygen-electrolyte RDF, for different pair potentials including PME. The results are for NaCl solutions with a salt concentration of 2.0 mol kg$^{-1}$, with $R_c=1.28$ nm. The inset shows a magnification to emphasize the discrepancies between pair potentials at long distances.}
    \label{fig:logOfErrSalt}
\end{figure}

Fig.~\ref{fig:logOfErrSalt} shows RDFs between water and salt for the different potentials. Similarly to the oxygen-oxygen RDFs (Fig.~\ref{fig:rdf_m1}), the lowest order of $P$, here $P=2$, yields results deviating the most compared to PME. Increasing the order up to $P=7$ improves the agreement, whereas $P=\infty$ again yields results showing increased deviation from PME.

\section{Discussion}

For the simulated water-system using $R_c=1.28$~nm we have found $P=\{4,5,6\}$ to closely match Ewald and PME results.
This agrees with results for the dipolar $q$-potential in Stockmayer systems using a similar cutoff\cite{Stenqvist2019shortranged}.
There is thus a limit on how many cancellations are appropriate. Though the discrepancies between the pair potentials and Ewald are oscillating, the peaks consistently occur around the cutoff distance. 
Arguably, the dipole-dipole correlation as a function of separation (Fig.~\ref{fig:mumu}) is more sensitive to electrostatic interactions than the radial distribution function (Fig.~\ref{fig:rdf_m1}).
For example, in previous studies using the Wolf formalism it has been shown that suboptimal combinations of cutoff and damping parameter can lead to unphysical results, where water dipoles align towards each other \cite{Stenqvist2015Direct,Wirnsberger2016}. We do not see such tendencies with the $q$-potential which for all conditions show the same trends as Ewald. The results of the dipole-dipole correlation difference to Ewald (inset of Fig.~\ref{fig:mumu}) at $R_c=1.28$~nm display distinguishable oscillating trends. However, the amplitudes are of the same magnitude as the noisy segments at other distances, in contrast with the results using $R_c=0.96$~nm. Ergo, this suggests $R_c=1.28$~nm to be the lowest cutoff to give Ewald-like results on every distance. 
%In contrast, the SP1 gives a lower density than Ewald, with a relative deviation similar to that between the $q$-potential and Ewald.
%SP3 gives a very similar density as Ewald and experimental values. 
%Some discrepancy between truncated potentials and Ewald are to be expected and in this case reflects stronger ($q$-potential) or weaker (SP1) electrostatic interactions compared to how the force field was originally developed.
Further, the discrepancies in Fig.~\ref{fig:rdf_m1} and Fig.~\ref{fig:mumu} seems to imply deviations from Ewald in densities, which is most clearly seen for the results using $R_c=0.96$~nm.

From a theoretical perspective we expect all long-ranged interactions to vanish by using $P=\infty$, \emph{i.e.} cancelling every moment locally.
From the point-of-view of long-ranged interactions this would solve the problem of summation \emph{if} indeed the surrounding medium did induce perfect moment cancellation.
In this study we observe that even water with its rather high relative permittivity does not act as a perfect conductor.
An infinite $P$ suppresses all local fluctuations and gives misrepresented short-ranged interactions. Thus, choosing $P$ too large results in over-damping of the system compared to the reference, which suggests that moment cancellation gives physical results only up to a point. This is nonetheless only the case for the two smaller cutoffs, while for the largest one there is no upper limit for the optimal $P$.
As $R_c\to\infty$ the $q$-potential collapses to the unmodified exact Coulomb potential which is independent of $P$.
Conversely, for $R_c\to 0$ no $P$-values give a proper description of the electrostatics and the optimal interval for $P$ varies with the cutoff.
Optimal values are about $P=5$ for $R_c=0.96$~nm; $P\in\{4,5,6\}$ for $R_c=1.28$~nm; and $P\in\{3,4,...\}$ for $R_c=1.6$~nm.
%Roughly, we have that $R_c=0.96$~nm gives $P=5$, $R_c=1.28$~nm gives $P\in\{4,5,6\}$, and $R_c=1.6$~nm gives $P\in\{3,4,...\}$.
Note that even the most precise results using $R_c=0.96$~nm are still inferior to a  mediocre parameterization at $R_c=1.6$~nm.
The optimal interval seemingly widen with increasing cutoff, and as discussed in Section~\ref{sec:gentheory} we find the expected lower value $P=3$ for a sufficiently large cutoff. 

The reason why all pair potentials reproduce PME activity derivatives may either be similarities in their RDFs or that the differences cancel while integrating to get the KBIs. For the q-potential, even the lowest order depicted, $P=2$, with RDFs deviating significantly more from PME relative to higher orders, yields accurate activity derivatives. This implies that the oscillating differences observed between RDFs using PME and the other pair potentials cancel when integrated, and do not necessarily pose an obstacle for retrieving accurate activity derivatives. 

As for the computational complexity, the Ewald summation real space pair potential is proportional to $\mathcal{N}\cdot R_c^3$, where the prefactor is implementation dependent.
Evaluation of the specifically used short-ranged function $\mathcal{S}(q)$ is part of this prefactor.
The complementary error-function, ${\rm erfc}$, used in Ewald is commonly approximated\cite{abramowitz1965handbook} using seven multiplications and one ${\rm exp}$ call where the latter requires at least ten times more clock cycles than a single multiplication.

The corresponding \emph{exact} evaluation of the $q$-potential short-ranged function $(q;q)_P$ includes $2(P-1)$ multiplications which for any relevant choice of $P \lesssim  5$ is significantly faster than the real space part of Ewald and PME.
In addition to the real space evaluation, Ewald and PME include a reciprocal space. The optimal real space cutoff for Ewald summation in the simulated bulk system can roughly be evaluated\cite{kolafa1992cutoff} to $R_c\approx 1.6$~nm.
We find that a lower cutoff can be used for the $q$-potential.
%to produce accurate results.
Therefore, for all valid $P$ values and $R_c\le 1.6$~nm we find that the $q$-potential is computationally less costly than both Ewald summation and PME, and scales with $P\cdot \mathcal{N}\cdot R_c^3$.

\section{Conclusion}\label{sec:conclusion}

We have presented a new theory for summation of long-ranged electrostatic interactions which relies on electrostatic moment and derivative cancellation, and has computational complexity of $\mathcal{O}(\mathcal{N})$.
The new method is a generalization of the Wolf method~\cite{wolf1999exact} with the advantages that it
(i) allows for cancellation of \emph{any} number of moments using an integer;
(ii) is free of operational damping parameters needed in similar algorithms;
(iii) is mathematically rigorous and has a simple form; and
(iv) reproduces results from Ewald and PME summation techniques for water and electrolyte solutions.

Results for solution structures, dielectric properties, diffusion coefficients, and activity derivatives of aqueous electrolyte systems agree well with Ewald and PME for a range of $P$ values, \emph{i.e.} number of cancelled moments.
We find that $R_c=1.28$~nm ($\sim 4$ molecular diameters) and $P=\{4,5,6\}$ gives appropriate results, where the interval size widens as the cutoff increases. The $q$-potential method is therefore a valid alternative to Ewald and PME at a lower computational cost.
The methodology has been expanded to dipolar interactions and is applicable also for explicit polarization simulations.

\acknowledgement
We thank LUNARC in Lund for computational resources, and for financial support:
S\"{o}dra Research foundation;
the Swedish Research Council;
and the Swedish Foundation for Strategic Research.
Finally, we thank the referees for valuable comments.

\suppinfo
The Supporting Information is available free of charge on the ACS Publications website at DOI:XYZ and includes: derivation of image charges, derivation of the self-energy, derivation of the dielectric constant expression, and simulation results of larger systems. 

The scripts used to produce the data for this work can be found at \href{https://github.com/mlund/SI-qpotential}{https://github.com/mlund/SI-qpotential}. An implementation of the derived potential as well as other truncation based schemes are also available in the CoulombGalore C++ package\cite{CoulombGalore}.

\appendix

\section{Kirkwood-Buff Theory}\label{app:F}
For the simulated $\mathcal{N}\mathcal{P}T$ ensemble, we considered the following running integral
\begin{equation}
\hat{G}_{AB}(R) = 4{\pi}\int_{0}^{R}[g_{AB}^{\mathcal{N}\mathcal{P}T}(r)-1]r^2\textrm{d}r.
\label{eq:apprKBint}
\end{equation}
where the hat indicates that RDFs in the $\mathcal{N}\mathcal{P}T$ ensemble is used in place of the ones in the grand canonical ensemble, and $R$ is the distance of truncation.
The excess coordination number of component $B$ around component $A$, \emph{i.e.} the excess number of particles $B$ around particles $A$ compared to bulk composition, is obtained as $\Delta N_{AB}=\rho_{B}G_{AB}$, where $\rho_B$ is the average number density of particles $B$ in the system. 

Due to the finite size of simulated systems, some complications arise when applying KB theory. Firstly, the calculated RDFs usually does not converge within the simulation cell. This is most easily understood considering the excess coordination number $\Delta N_{AB}$. In a finite system containing non-ideal particles, the excess coordination numbers between the particles are non-zero due to interactions. Since the system is finite, an excess or deficit of for instance particles $B$ around particles $A$ will cause a corresponding deficit or excess of particles $B$ in the bulk. Hence the bulk composition, $\rho_{AB}^{(bulk)}$, will be different from $\rho_B$ which is used for normalization according to $g_{AB}(r)=\rho_{AB}(r)/\rho_B$. To make the RDFs asymptotically approach unity, we used the following correction factor \cite{Hess2009, Ganguly2013} which scales the RDF so that it is normalized with $\rho_{AB}^{(bulk)}$ rather than $\rho_B$ 

\begin{equation}
C_{AB}^{(1)}(r) = \frac {N_B\left (1-\frac{V(r)}{V_{cell}} \right )} {{N_B\left (1-\frac{V(r)}{V_{cell}} \right )}-\Delta N_{AB}(r)-\delta _{AB}}.
\label{eq:1stCorr}
\end{equation}
Here $N_B$ is the total number of particles $B$, $V(r)$ is the volume of a sphere with radius $r$, $V_{cell}$ is the volume of the simulation cell, and $\delta_{AB}$ is the Kronecker delta. In the expression, the numerator is the number of particles $B$ that would occupy the volume beyond $r$ if the composition would be uniform, while the denominator approximates the true number of particles $B$ in the same volume. 

\begin{equation}
G_{AB}=\frac{1}{V}\int_{V}\int_{V}[g_{AB}^{\mu VT}(r_{12})-1]\textrm{d}\mathbf{r_1}\textrm{d}\mathbf{r_2}
\label{eq:defKBint}
\end{equation}
Secondly, for closed systems one cannot reduce the double integral in Eq.~\ref{eq:defKBint} to the single integral in Eq.~\ref{eq:defKBint_red} since the integration domain of $r$ is no longer independent of $\mathbf{r_1}$.
To account for this, a method has been suggested\cite{Krger2013} where the exact KBI is found by applying a geometry dependent factor to the single integral and evaluating the expression in the thermodynamic limit, $1/R \rightarrow 0$.
Alternatively~\cite{Krger2013}, it has been suggested that an extrapolated expression of the exact KBI can be obtained by applying the following factor directly in the single integral, which is the method we used,
\begin{equation}
C_{AB}^{(2)}(r) = 1-\left( \frac{r}{R} \right)^3.
\label{eq:2ndCorr}
\end{equation}

To obtain approximate KBIs accounting for the system being both closed and of finite size, we simultaneously apply the two corrections\cite{Milzetti2018} according to
\begin{equation}
G_{AB}\approx  \hat{G}_{AB}^*(R)=4\pi\int_{0}^{R} [g_{AB}^{\mathcal{N}pT}(r)C_{AB}^{(1)}(r)-1]r^{2}C_{AB}^{(2)}(r)\textrm{d}r
\label{eq:dblApprKBint}
\end{equation}
where the star indicates that correction factors are applied. For the activity derivatives, the integral was then evaluated at a value of $R$ where convergence is obtained, corresponding to the distance from the center particle at which bulk composition is reached.

The experimental activity coefficients\cite{Robinson1959} were used to find the values of the parameters $b_j$ in the following fitting function\cite{Gee2011}, relating the activity coefficient to the concentration on the molal scale,
\begin{equation}
\ln \gamma(m) = -\frac{1.18\sqrt{m}}{1+b_1\sqrt{m}} - \ln(1-b_2 m) + b_3 m + b_4 m^2.
\label{eq:fit_g}
\end{equation}
The obtained values of $b_j$ are reported in Table~\ref{tab:S1}.

\begin{table}[ht]
\caption{Fitting parameters used to obtain activity coefficients as a function of salt molality using Equation \ref{eq:fit_g}.}
\begin{tabular}{lrrrr}\hline
          Salt & $b_1$ & $b_2$ & $b_3$ & $b_4$ \\\hline
NaCl   & 1.4369 & 0.0054 & 0.0495 & 0.0092 \\
NaI    & 1.4681 & 0.1361 & 0.0344 &-0.0102 \\
\hline
\label{tab:S1}
\end{tabular}
\end{table}
Since the experimental activity coefficients used in this study are on the molal scale, we converted the concentration scale to be in the unit of number density in order to calculate the activity derivatives as $a_{c}'=\left ( \partial \ln a_c / \partial \ln \rho_c \right )_{\mathcal{P}T}$. For the conversion, we used experimental density data\cite{Zaytsev1992,Lalibert2004}.

\begin{table}[ht]
\caption{Lennard-Jones parameters for Na$^{+}$, Cl$^{+}$, and I$^{-}$: charge $z$, diameter $d$, and interaction-strength $\epsilon$. To the right the number of water-molecules in each simulation is presented for the different molal salt concentrations (mol$^{-1}$ kg$^{-1}$).}
\begin{tabular}{lllllllll}\hline
\multirow{2}{*}{Salt} & \multirow{2}{*}{Ion} & \multirow{2}{*}{$z$} & \multirow{2}{*}{$d$/nm} & \multirow{2}{*}{$\epsilon$/kJ$\cdot$mol$^{-1}$} & & $\mathcal{N}_{H_2O}$ & & \\  & & & & & 1.0 & 2.0 & 3.0 & \\\hline
\multirow{2}{*}{NaCl} & Na$^+$ & +1.0 & 0.255 & 0.280 & \multirow{2}{*}{5963} & \multirow{2}{*}{5252} & \multirow{2}{*}{4672} \\
& Cl$^-$ & -1.0 & 0.440 & 0.418 &  & & \\
\multirow{2}{*}{NaI} & Na$^+$ & +1.0 & 0.255 & 0.280 & \multirow{2}{*}{5796} & \multirow{2}{*}{5026} & \multirow{2}{*}{4402}  \\ 
& I$^-$ & -1.0 & 0.491 & 0.158 &  & & \\\hline
\label{tab:S2}
\end{tabular}
\end{table}

\bibliography{manuscript}

\end{document}

\section{Derivation of image charges}\label{app:B}
By using $r_p=c_pr$ and $\hat{z}_{p} = z_{p}/z$ it is possible to transform Eq.~7 (main text) into Eq.~\ref{eq:transformed}.
\begin{equation}
\label{eq:transformed}
\begin{bmatrix}
       1    \\[0.3em]
       1  \\[0.3em]
       \vdots          \\[0.3em]
       1      
     \end{bmatrix} + \begin{bmatrix}  
  1 & 1 & \cdots & 1 \\
  c_1 & c_2 & \cdots & c_P \\
  \vdots  & \vdots  & \ddots & \vdots  \\
  c_1^{P-1} & c_2^{P-1} & \cdots & c_P^{P-1}
     \end{bmatrix}
     \begin{bmatrix}
       \hat{z}_{1}    \\[0.3em]
       \hat{z}_{2}  \\[0.3em]
       \vdots          \\[0.3em]
       \hat{z}_{P} 
     \end{bmatrix}
     = \begin{bmatrix}
       0    \\[0.3em]
       0  \\[0.3em]
       \vdots          \\[0.3em]
       0      
     \end{bmatrix}.
\end{equation}
The solution\cite{el2003explicit} to Eq.~\ref{eq:transformed}, by using $c_p=q^{-p}$, is
\begin{equation}
\label{eq:moment_derivation_0}
\hat{z}_p = -\frac{   \prod_{\substack{i = 1 \\ i \ne p }}^{P} (1 - q^{-i})   }{\prod_{\substack{i = 1 \\ i \ne p }}^{P} (q^{-p} - q^{-i})  }.
\end{equation}
In the bottom product of Eq.~\ref{eq:moment_derivation_0} we factor out $q^{-p}$, and then split all products into cases when $i<p$ and $i>p$. These modifications are shown in Eq.~\ref{eq:moment_derivation_1}.
\begin{equation}
\label{eq:moment_derivation_1}
\hat{z}_p  = -\frac{   \prod_{\substack{i = 1 \\ i \ne p }}^{P} (1 - q^{-i})   }{\prod_{\substack{i = 1 \\ i \ne p }}^{P} q^{-p}(1 - q^{-i+p})  } = \frac{   \prod_{i = 1}^{p-1} (1 - q^{-i})\prod_{i = p+1}^{P} (1 - q^{-i})   }{q^{-p(P-1)}\prod_{i = 1 }^{p-1} (1 - q^{-i+p})\prod_{i = p+1}^{P} (1 - q^{-i+p})  }
\end{equation}
Further modification by variable substitution ($i^{\prime}=i-p$ in the top and bottom right products, and $i^{\prime}=-i+p$ in the bottom left product) gives Eq.~\ref{eq:moment_derivation_2}.
\begin{equation}
\label{eq:moment_derivation_2}
\hat{z}_p  = -\frac{   \prod_{i = 1}^{p-1} (1 - q^{-i})\prod_{i = 1}^{P-p} (1 - q^{-i-p})   }{q^{-p(P-1)}\prod_{i = 1 }^{p-1} (1 - q^{i})\prod_{i = 1}^{P-p} (1 - q^{-i})  }
\end{equation}
From Eq.~\ref{eq:moment_derivation_2} and onward we will re-letter the new index symbol $i^{\prime}$ with the old $i$ when using variable substitution, i.e. $i^{\prime}\to i$ in this case. This to avoid multiple indexes and thus confusions with other entities. Further simplification of the two left products (by factoring out $q^{-i}$ from the top one) gives Eq.~\ref{eq:moment_derivation_2p}.
\begin{equation}
\label{eq:moment_derivation_2p}
\hat{z}_p =  (-1)^pq^{p(2P-p-1)/2} \frac{   \prod_{i = 1}^{P-p} (1 - q^{-i-p})   }{\prod_{i = 1}^{P-p} (1 - q^{-i})  }
\end{equation}
We now factor out $q^{-i-p}$ from the top product and $q^{-i}$ from the bottom product giving Eq.~\ref{eq:next}, where we also note the cancellation of the $(-1)^{P-p}$ factors.
\begin{equation}
\label{eq:next}
\hat{z}_p =  (-1)^pq^{p(2P-p-1)/2} \frac{\prod_{i = 1}^{P-p}q^{-i-p}  \prod_{i = 1}^{P-p} (1 - q^{i+p})   }{\prod_{i = 1}^{P-p}q^{-i}\prod_{i = 1}^{P-p} (1 - q^{i})  }
\end{equation}
Simplification of the left products yield
\begin{equation}
\label{eq:nextnext}
\hat{z}_p = (-1)^pq^{p(p-1)/2} \frac{\prod_{i = 1}^{P-p} (1 - q^{i+p})   }{\prod_{i = 1}^{P-p} (1 - q^{i})  }.
\end{equation}
By making the variable substitution $i^{\prime} = P-i-p+1$ in the top product we get Eq.~\ref{eq:nextnextnext} where we note that the products together are equal to the $q$-binomial coefficient\cite{comtetadvanced} ${P \brack P-p}_q = {P \brack p}_q$.
\begin{equation}
\label{eq:nextnextnext}
\hat{z}_p = (-1)^pq^{p(p-1)/2} \frac{\prod_{i = 1}^{P-p} (1 - q^{P-i+1})   }{\prod_{i = 1}^{P-p} (1 - q^{i})  }
\end{equation}
Thus we now have arrived at
\begin{equation}
\label{eq:moment_derivation_4}
\hat{z}_p  = (-1)^pq^{p(p-1)/2}{P \brack p}_q.
\end{equation}

\section{Self-energy}\label{app:D}
Starting from Eq.~7 (main text) we note that if a particle is positioned in the origin, i.e. $r=0$, then the equation becomes
\begin{equation}
\label{eq:finalselfMatrix}
\begin{bmatrix}
       1    \\[0.3em]
       0  \\[0.3em]
       \vdots          \\[0.3em]
       0      
     \end{bmatrix}z + \begin{bmatrix}  
  1 & 1 & \cdots & 1 \\
  r^{\prime}_1 & r^{\prime}_2 & \cdots & r^{\prime}_P \\
  \vdots  & \vdots  & \ddots & \vdots  \\
  r^{\prime P-1}_1 & r^{\prime P-1}_2 & \cdots & r^{\prime P-1}_P
     \end{bmatrix}
     \begin{bmatrix}
       z_{1}^{\prime}    \\[0.3em]
       z_{2}^{\prime}  \\[0.3em]
       \vdots          \\[0.3em]
       z_{P}^{\prime} 
     \end{bmatrix}
     = \begin{bmatrix}
       0    \\[0.3em]
       0  \\[0.3em]
       \vdots          \\[0.3em]
       0      
     \end{bmatrix}.
\end{equation}
Here we have indexed the image charges with primes as to distinguish them from the charges when we calculate the potential from a particle at position $r>0$. Note that in Eq.~\ref{eq:finalselfMatrix} there is only a charge present due to the centered particle and no higher order moments. Assuming that the image particles needed to cancel this charge (and all higher order moments generated by themselves in the process) are positioned at $r^{\prime}_p=c^{\prime}_pr^{\prime}$ where $r^{\prime}>0$ is any point, then Eq.~\ref{eq:finalselfMatrix} converts to Eq.~\ref{eq:finalMatrix_self} which has its solution\cite{el2003explicit} shown in Eq.~\ref{eq:moment_derivation_0_self} where we have used $c^{\prime}_p=q^{-p}$.
\begin{dmath}
\label{eq:finalMatrix_self}
\begin{bmatrix}
       1    \\[0.3em]
       0  \\[0.3em]
       \vdots          \\[0.3em]
       0      
     \end{bmatrix} + \begin{bmatrix}  
  1 & 1 & \cdots & 1 \\
  c^{\prime}_1 & c^{\prime}_2 & \cdots & c^{\prime}_P \\
  \vdots  & \vdots  & \ddots & \vdots  \\
  c^{\prime P-1}_1 & c^{\prime P-1}_2 & \cdots & c^{\prime P-1}_P
     \end{bmatrix}
     \begin{bmatrix}
       \hat{z}^{\prime}_{1}    \\[0.3em]
       \hat{z}^{\prime}_{2}  \\[0.3em]
       \vdots          \\[0.3em]
       \hat{z}^{\prime}_{P} 
     \end{bmatrix}
     = \begin{bmatrix}
       0    \\[0.3em]
       0  \\[0.3em]
       \vdots          \\[0.3em]
       0      
     \end{bmatrix}
\end{dmath}
\begin{equation}
\label{eq:moment_derivation_0_self}
\hat{z}^{\prime}_p = -\frac{\prod_{\substack{i = 1 \\ i \ne p }}^{P} (-q^{-i})}{\prod_{\substack{i = 1 \\ i \ne p }}^{P} (q^{-p} - q^{-i})  }
\end{equation}
By condensing these products into one, and splitting the result as to give products for $i<p$ and $i>p$, we get
\begin{equation}
\label{eq:moment_derivation_0_self2}
\hat{z}^{\prime}_p = -\frac{1}{\prod{\substack{i = 1 \\ i \ne p }}^{P}(1-q^{-p+i})} = -\frac{1}{\prod_{\substack{i = 1}}^{p-1}(1-q^{-p+i})\prod_{\substack{i = p+1 }}^{P}(1-q^{-p+i})}.
\end{equation}
Variable substitution using $i^{\prime}=i-p$ in the right product gives
\begin{equation}
\label{eq:moment_derivation_0_self3}
\hat{z}^{\prime}_p = -\frac{1}{\prod_{\substack{i = 1}}^{p-1}(1-q^{-p+i})\prod_{\substack{i = 1 }}^{P-p}(1-q^{i})}
\end{equation}
and by factoring out $q^{-p+i}$ from the left product we get
\begin{equation}
\label{eq:moment_derivation_0_self4}
\hat{z}^{\prime}_p = -(-1)^{p-1}\frac{q^{(p-1)p/2}}{\prod_{\substack{i = 1}}^{p-1}(1-q^{p-i})\prod_{\substack{i = 1 }}^{P-p}(1-q^{i})}.
\end{equation}
Using these moments, the self-energy becomes
\begin{equation}
\label{eq:self_1}
E_{{\rm Self}} = \frac{e^2}{4\pi\varepsilon_0\varepsilon_r}\sum_{j = 1}^{N} \frac{z_j^2}{R_c}\sum_{p=1}^P \frac{\left(-(-1)^{p-1}\frac{q^{(p-1)p/2}}{\prod_{\substack{i = 1}}^{p-1}(1-q^{p-i})\prod_{\substack{i = 1 }}^{P-p}(1-q^{i})} \right)}{q^{-(p-1)}}.
\end{equation}
Reshuffling the terms in Eq.~\ref{eq:self_1} gives Eq.~\ref{eq:self_2}.
\begin{equation}
\label{eq:self_2}
E_{{\rm Self}} = -\frac{e^2}{4\pi\varepsilon_0\varepsilon_r}\sum_{j = 1}^{N} \frac{z_j^2}{R_c}\sum_{p=1}^P(-1)^{p-1}\frac{q^{(p-1)p/2}q^{(p-1)}}{\prod_{\substack{i = 1}}^{p-1}(1-q^{p-i})\prod_{\substack{i = 1 }}^{P-p}(1-q^{i})}
\end{equation}
The denominators in Eq.~\ref{eq:self_2} are polynomials with only non-negative powers. Thus, if $q\to 0$ only the constant term $1$ will be none-vanishing. In the same limit the numerator will be zero for every $p>1$ and thus the entire far-right sum will equal one in the limit $q\to 0$, which stems from the $p=1$ term. The final expression (independent of $P$) for the far right sum in the limit $q\to 0$ is thus $1$ as is shown in Eq.~\ref{eq:self_3}. The choice of $q$ seems somewhat arbitrary however our choice of $q\to 0$ comes from the following arguments: In the original derivation for the potential we chose to mirror the particle position in the cut-off as to get the image particle positions. However, this is not possible to do when the particle is in the origin (since then we would have to divide by zero). Thus, we choose to mirror an identical particle infinitesimally close ($q\to 0$) to the origin.
\begin{equation}
\label{eq:self_3}
E_{{\rm Self}} = -\frac{e^2}{4\pi\varepsilon_0\varepsilon_rR_c}\sum_{j = 1}^{N} z_j^2
\end{equation}

\section{Dielectric constant}\label{app:H}
The dielectric constant, $\varepsilon_r$, has been derived within a known theoretical framework\cite{neumann1986computer} where the key equation is
\begin{dmath}
\label{eq:orgDiel}
\frac{\varepsilon_r - 1}{\varepsilon_r + 2}\left[1 - \frac{\varepsilon_r - 1}{\varepsilon_r + 2}\tilde{T}(0) \right]^{-1} = \frac{1}{3\varepsilon_0}\frac{\langle M^2 \rangle}{3Vk_BT}.
\end{dmath}
Here $\langle M^2\rangle$ are the fluctuations of the dipole moment $\boldsymbol{M}=\sum_{i=1}^{\mathcal{N}}\boldsymbol{\mu}_i$, $k_B$ is the Boltzmann constant, and $V$ the volume of the unit cell. Different values of $\tilde{T}(0)$ is used depending on the method. In the following derivations we want to stress that $q=q(r)$. In order to get first higher order interactions, we write
\begin{equation}
\nabla\left(\frac{\mathcal{S}(q)}{r} \right) = \frac{\nabla\mathcal{S}(q)}{r} + \mathcal{S}(q)\nabla\left(\frac{1}{r}\right)
\end{equation}
where $\nabla$ is the gradient operator. Further second higher order interactions then are obtained as
\begin{equation}
\nabla^T\nabla\left(\frac{\mathcal{S}(q)}{r} \right)  =  \frac{\nabla^T\nabla\mathcal{S}(q)}{r} + \nabla^T\mathcal{S}(q)\nabla\left(\frac{1}{r}\right) + \nabla^T\left(\frac{1}{r}\right)\nabla\mathcal{S}(q) + \mathcal{S}(q)\nabla^T\nabla\left(\frac{1}{r}\right).
\end{equation}
Note that $\mathcal{S}(q)$ is not angle-dependent and thus
\begin{equation}
\nabla\mathcal{S}(q) = {\bf \hat{r}}\frac{\partial}{\partial r}\mathcal{S}(q).
\end{equation}
However, by further apply $\nabla^T$ to this we get
\begin{equation}
\nabla^T\nabla\mathcal{S}(q) = {\bf \hat{r}}^T{\bf \hat{r}}\frac{\partial^2}{\partial r^2}\mathcal{S}(q) + \frac{\frac{\partial}{\partial r}\mathcal{S}(q)}{r}\left({\bf I} - {\bf \hat{r}}^T{\bf \hat{r}}   \right).
\end{equation}
The total expression for $\nabla^T\nabla\left(\mathcal{S}(q)/r \right)$ can be parted like
\begin{equation}
\label{eq:T2_exp}
\nabla^T\nabla\left(\frac{\mathcal{S}(q)}{r} \right) = a(r)\left(3\bf{\hat{r}}^T{\bf \hat{r}} - {\bf I}\right) + b(r){\bf I}
\end{equation}
where
\begin{equation}
\label{eq:T2_exp_a}
a(r) = \frac{\frac{\partial^2}{\partial r^2}\mathcal{S}(q)}{3r} - \frac{\frac{\partial}{\partial r}\mathcal{S}(q)}{r^2} + \frac{\mathcal{S}(q)}{r^3}
\end{equation}
and
\begin{equation}
\label{eq:T2_exp_b}
b(r) = \frac{\frac{\partial^2}{\partial r^2}\mathcal{S}(q)}{3r}.
\end{equation}
In order to get the proper evaluation of the dielectric constant we have to evaluate the integrals\cite{neumann1986computer}
\begin{equation}
A(k) = -3\int_{0}^{\infty}r^2j_2(kr)a(r)dr
\end{equation}
and
\begin{equation}
B(k) = 3\int_{0}^{\infty}r^2j_0(kr)b(r)dr.
\end{equation}
For $k=0$, i.e. evaluation of the static dielectric constant, the spherical Bessel functions becomes $j_0(0) = 1$ and $j_2(0) = 0$. Therefore $A(0)$ has the trivial solution zero (since the singularity of $a(r)$ in $r=0$ has been explicitly dealt with\cite{neumann1986computer}) and we now only have to evaluate $B(0)$. Thus, we have
\begin{equation}
B(0) = \int_{0}^{R_c}r\frac{\partial^2}{\partial r^2}\mathcal{S}(q)dr
\end{equation}
where the limit $\infty$ has changed to $R_c$ due to the fact that we use $\mathcal{S}(q) \equiv 0$ for $q>1$, that is $r> R_c$. Integration by parts gives
\begin{equation}
B(0) = \left[r\frac{\partial}{\partial r}\mathcal{S}(q) \right]_{0}^{R_c} - \int_0^{R_c}\frac{\partial}{\partial r}\mathcal{S}(q)dr = 1
\end{equation}
which is true for all the tested pair potentials except $q$-potential using $P=1$ where $B(0)=0$. Finally, we note that $\tilde{T}(0) = B(0)$ and thus the derivation is done. 

\section{Larger systems}
The density, dielectric constant, Kirkwood factor $G_K$, standard deviation of the total energy, and diffusion coefficient, for the different potentials are presented in Table~\ref{tbl:denteSI}. Here $*$ indicate a large system, i.e. a system where $R_c$ is less than a fourth of the cubic system side-length. In this case the number of water molecules was $\mathcal{N}=5000$. The three consistent differences we see are a larger diffusion coefficient, $G_K$, and $\sigma_E$ in a larger system. System dependencies of the first are well-known\cite{Tazi_2012} and our results are consistent with this observation. Increases of the second are small and could fall under the uncertainty given by the standard deviation. Yet the increase of the standard deviation of the energy is noticeable for all $q$-potentials and Ewald. However both SP1 and SP3 are consistently low in this regard.

\begin{table}[t]
    \centering
    \begin{tabular}{ c | c c c | c c | c c c | c c | c c c | c c |}
    %\hline
    \multirow{2}{*}{Potential} & \multicolumn{5}{c|}{$R_c=1.28$~nm, $*$} & \multicolumn{5}{c|}{$R_c=1.28$~nm} & \multicolumn{5}{c|}{$R_c=1.60$~nm} \\ & $\rho$ & $\varepsilon_r$ & $G_K$ & $\sigma_E$ & $D$ & $\rho$ & $\varepsilon_r$ & $G_K$ & $\sigma_E$ & $D$ & $\rho$ & $\varepsilon_r$ & $G_K$ & $\sigma_E$ & $D$ \\\hline
    $q(P=1)$ & 1099 & --- & 3.5 & --- & --- & 1106 & ---  & 3.2 & --- & --- & 1073 & --- & 3.0 & --- & --- \\
    $q(P=2)$ & 1002 & 68 & 3.0 & 27 & 2.3 & 1002 & 75 & 3.0 & 15 & 2.2 & 1000 & 69 & 2.9 & 11 & 2.7 \\
    $q(P=3)$ & 1001 & 68 & 3.0 & 11 & 2.5 & 1001 & 70 & 3.0 & 8 & 2.5 & 1000 & 76 & 3.0 & 3 & 2.4 \\
    $q(P=4)$ & 1001 & 67 & 3.0 & 17 & 2.6 & 1000 & 67 & 2.9 & 6 & 2.5 & 1000 & 69 & 2.9 & 10 & 2.4 \\
    $q(P=5)$ & 1001 & 67 & 3.0 & 18 & 2.6 & 1001 & 67 & 2.9 & 8 & 2.6 & 1000 & 72 & 2.9 & 12 & 2.3 \\
    $q(P=6)$ & 1001 & 69 & 3.1 & 11 & 2.7 & 1001 & 71 & 3.0 & 4 & 2.4 & 1000 & 68 & 2.9 & 3 & 2.7 \\
    $q(P=7)$ & 1001 & 66 & 3.0 & 12 & 2.8 & 1001 & 68 & 2.9 & 4 & 2.4 & 1000 & 69 & 2.9 & 3 & 2.5 \\
    $q(P=8)$ & 1001 & 69 & 3.0 & 22 & 2.7 & 1001 & 70 & 2.9 & 6 & 2.6 & 1000 & 71 & 3.0 & 6 & 2.5 \\
    $q(P=\infty)$ & 1001 & 76 & 3.1 & 18 & 2.7 & 1001 & 66 & 2.9 & 4 & 2.7 & 1000 & 67 & 2.9 & 4 & 2.5 \\\hline
    SP1 &   996 & 74 & 3.1 & 1 & 3.1 &   996 & 69 & 2.9 & 1 & 2.8 &   996 & 68 & 2.9 & 1 & 2.8\\
    SP3 &   998 & 74 & 3.0 & 0 & 2.7 &   998 & 66 & 2.9 & 0 & 2.5 &   998 & 71 & 3.0 & 0 & 2.5 \\\hline
    Ewald & 998 & 68 & 3.0 & 17 & 2.8 & 998 & 73 & 3.0 & 2 & 2.6 & 998 & 69 & 3.0 & 6 & 2.9 \\
    \hline
    Exp. & 997 & 79 & --- & --- & 2.3 & 997 & 79 & --- & --- & 2.3 & 997 & 79 & --- & --- & 2.3 \\
    %\hline
    \end{tabular}
    \caption{Density $\rho$ [kg/m$^3$], relative dielectric constant $\varepsilon_r$ [unitless], Kirkwood factor $G_K$ [unitless], standard deviation of total energy $\sigma_E$ [kJ/mol], and diffusion coefficient $D$ [m$^2$/s/10$^{-9}$], for the different potentials applied on a bulk water-system and experimental reference\cite{harned1958physical,Mills1973Self}.}
    \label{tbl:denteSI}
\end{table}

\bibliography{manuscript}